\documentclass[prm,twocolumn, amsmath,amssymb,
reprint,%
author-year,%
author-numerical,%
dvipdfmx,
superscriptaddress
]{revtex4-1}

\usepackage{graphicx}
\usepackage{bm}
\usepackage{color}

\begin{document}

\title{Fluorite-type materials in the monolayer limit}

\author{Shota Ono}
\email{shota.ono.d3@tohoku.ac.jp}
\affiliation{Institute for Materials Research, Tohoku University, Sendai 980-8577, Japan}
\author{Ravinder Pawar}
\affiliation{Department of Chemistry, National Institute of Technology (NIT) Warangal, Warangal, Telangana, 506004, India}

\begin{abstract}
The 2H, 1T, and their distorted structures are known as prototype structures of $AB_2$ monolayers. Here, we study a puckered structure that is truncated from the (110) surface of fluorite-type materials. 53 fluorite-type materials are investigated based on first-principles approach. The formation energy calculations indicate that seven systems form the puckered structure in the monolayer limit, while other systems form either 1T, 2H, or distorted 1T structures. The puckered structures of PbF$_2$, PRh$_2$, and Ga$_2$Au exhibit negative Poisson's ratio (NPR) in the out-of-plane direction. An analytical model for the NPR is derived. The surface energy calculations predict the appearance of NPR. 
\end{abstract}

\maketitle

\section{Introduction}
Two-dimensional (2D) transition metal dichalcogenides have attracted attention due to their electronic, optical, and mechanical properties \cite{choi2017}. They are classed as $AB_2$ materials, where $A$ is a transition metal and $B$ is a chalcogen atom (S, Se, and Te), and exfoliated from 3D bulk due to their layered structure. 
They usually adopt either 2H or 1T structures and exhibit a structure-dependent property. 
For example, 2H MoS$_2$ is a semiconductor with a direct band gap, whereas 1T MoS$_2$ is a metal and shows a ferromagnetism under a tensile strain \cite{magneMoS22018}.
Distorted 1T structures have been studied in several $AB_2$ monolayers. 
The 1T$^\prime$ structure is a Peierls distorted phase, and 1T$^\prime$ WTe$_2$ shows a ferroelasticity that originates from the three equivalent directions of the distortion \cite{lili2016}. 
More complex geometries, such as 1T$^{\prime\prime}$ and 1T$^{\prime\prime\prime}$ structures, have also been investigated \cite{zhao2018}.
By performing high-throughput density-functional theory (DFT) calculations and assuming these structures, several structure maps have been proposed for $AB_2$ monolayers \cite{fukuda2021,silva2022,kumar2022}.

It is of fundamental importance to explore other structures different from 1T and related-structures.
Recently, 2D materials created from non-layered materials have also been synthesized experimentally \cite{ji2019,puthirath2021,balan2022}.
In addition, a wide variety of 2D materials has been predicted by cutting surfaces of non-layered materials.
Lucking {\it et al}. have predicted that an ultrathin layer truncated from the (111) surface of zincblend-type semiconductors relaxes to double-layered honeycomb structure, and exhibits topological electronic band structure \cite{2Dtraditional}. Friedrich {\it et al}. have studied electronic, optical, and magnetic properties of hematite and ilmenite in the monolayer limit \cite{friedrich2022}. The present authors have also studied structural and physical properties of 2D metallic systems \cite{ono2020,ono2020_Po,sangolkar2022}. Many elemental metals have hexagonal structures that are truncated from the (111) surface of the face-centered cubic structure or the (0001) surface of the hexagonal close-packed structure. 

\begin{figure}[b]
\center
\includegraphics[scale=0.35]{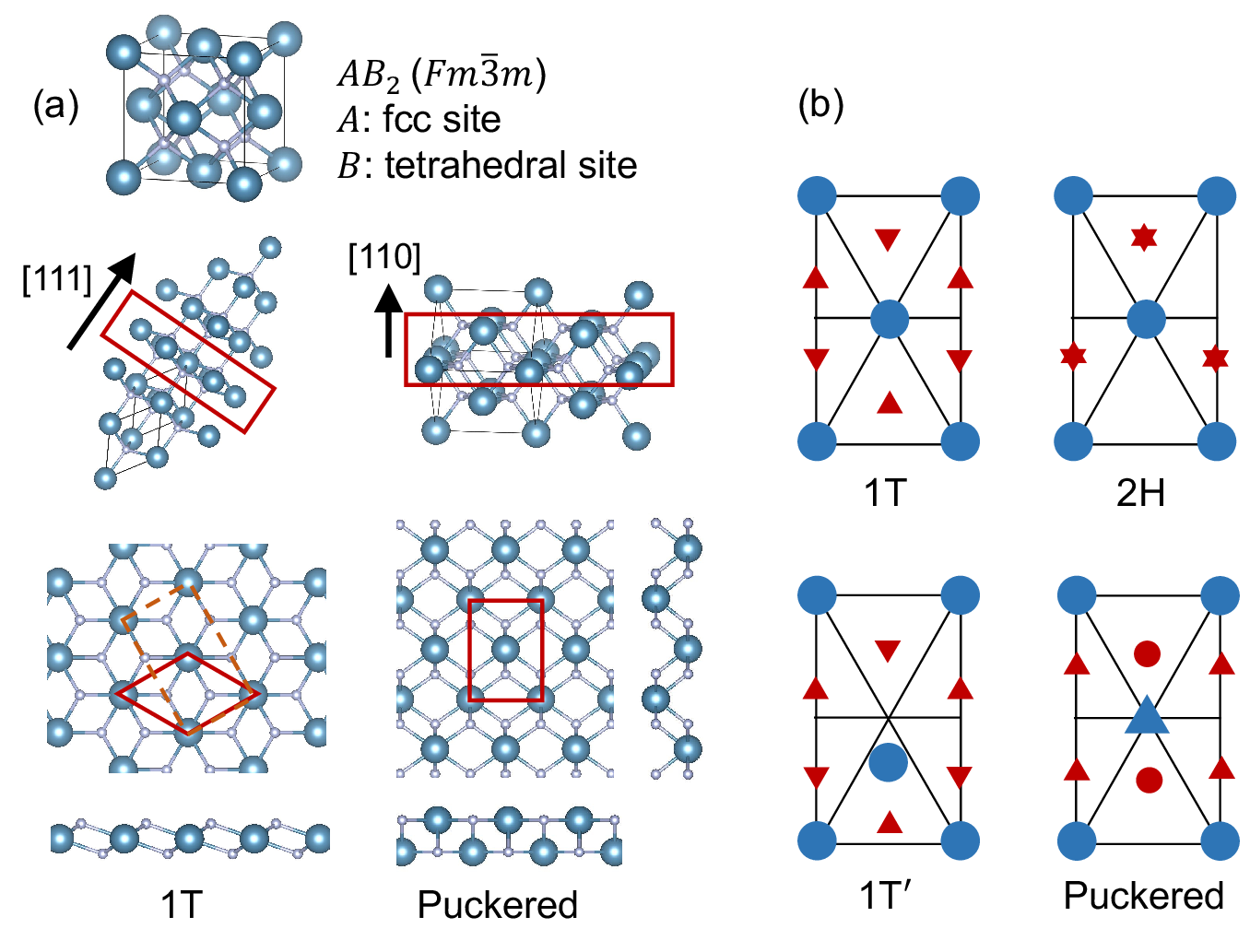}
\caption{(a) Crystal structure of fluorite-type material in the bulk form and the monolayer limit. 1T and puckered structures are truncated from the (111) and (110) surfaces, respectively. \texttt{VESTA} \cite{vesta} is used to visualize crystal structures. (b) Schematic illustration of rectangular-shaped unit cell for 1T, 2H, 1T$^{\prime}$, and PCK structures. Blue and red shapes indicate $A$ and $B$ atoms, respectively. Upward (downward) triangle is located above (below) $z=0$ plane, and a circle is located at $z=0$. } \label{fig1} 
\end{figure}

Fluorite-type (CaF$_2$-type) materials exhibit a cubic structure with a stoichiometry of $AB_2$, where $A$ atoms form a face-centered cubic structure and $B$ atoms occupy the tetrahedral sites. 
Therefore, two monolayer structures are truncated from their surface with keeping the stoichiometry of bulk (see Fig.~\ref{fig1}(a)). 
One is the 1T structure that is truncated from the (111) surface. 
The other monolayer is truncated from the (110) surface and exhibits a highly puckered (PCK) structure.
Although the PCK structure has been investigated theoretically for Be$_2$C \cite{naseri2019} and ZrS$_2$ \cite{abutalib2019}, such a structure is not the ground state. 
In addition, physical properties intrinsic to the PCK structure have not been explored yet. 

In this paper, we investigate the structural and physical properties of 53 fluorite-type materials in the monolayer limit by using first-principles approach.
Seven monolayers are identified to have the PCK structure, and among them, PbF$_2$, Ga$_2$Au, and PRh$_2$ monolayers have an out-of-plane negative Poisson's ratio (NPR), that is, their thickness increases when stretched.
PbF$_2$ and Ga$_2$Au monolayers are dynamically and thermodynamically stable, but PRh$_2$ monolayer is unstable at room temperature. 
By developing an analytical model, we demonstrate that the NPR is inherently present in the PCK structure. 
We also show that a linear relationship holds between the (111) and (110) surface energies, and PbF$_2$, Ga$_2$Au, and PRh$_2$ exhibit an anomalously small (110) surface energy. 

It is known that many puckered structures exhibit a negative Poisson's ratio in the out-of-plane direction. For example, phosphorene ($-0.027$) \cite{jiang2014,du2016}, arsenic ($-0.09$) \cite{han2015}, GeS ($-0.14$ within GGA-PBE and $-0.19$ with van der Waals correction) \cite{gomes2015}, SnSe ($-0.17$) \cite{zhang2016}, TiN ($-0.102$) \cite{zhou2017}, Ag$_2$S ($-0.52$) \cite{peng2019}, and SnS$_2$ ($-1.79$) \cite{wang2023}, where the value in a parenthesis indicates the Poisson's ratio $\nu$ at equilibrium condition. The Ga$_2$Au ($-0.6$) and PbF$_2$ ($-0.4$) monolayers studied in the present work also serve as 2D auxetic materials.

\section{Computational details}
\subsection{2D structures}
By using Materials Project database \cite{materalsproject} and pymatgen code \cite{pymatgen}, we first extracted 94 fluorite-type materials (spacegroup of $Fm\bar{3}m$). These materials have been found in inorganic crystal structure database (ICSD).  
We excluded $f$-electron systems (lanthanide and actinide compounds), some of hydrogen compounds (Li$_2$NH, K$_2$PtH$_4$, Ca$_2$RhH$_{5.4}$, and Sr$_2$RhH$_5$), high-temperature phase materials (Cu$_2$S, Cu$_2$Se, and Al$_2$O), and oxides (PbO$_2$ and BiO$_2$). H$_2$S and H$_2$Se were also excluded because of a large discrepancy between experimental and calculated lattice constants. Zirconia (ZrO$_2$) and hafnia (HfO$_2$) in the fluorite-type structure are known to be stable above 2650 K and 2870 K, respectively \cite{shin2018zirconia}, but these are studied in the present work. 

We study 53 fluorite-type materials that consist of $AB_2$ (19), $A_2B$ (14), and others (20), where $A$ is a metallic element and $B$ is H, F, Cl, O, S, Se, and Te. The $AB_2$-type system includes hydrides (8), fluorides (7), chlorides (2), and oxides (2). The $A_2B$ materials consist of alkali metals $A=$ Li (4), Na (4), K (3), and Rb (3) and chalcogen atoms $B=$ O, S, Se, and Te. The other materials consist of alkali earth metals Be (2) and Mg (4) and other metallic elements Rh (2), Ir (2), Pt (4), Au (3), Ni (1), Co (1), and Pd (1). 

We consider four crystal structures in the monolayer limit: 1T, 2H, 1T$^{\prime}$, and PCK structures. These are illustrated in Fig.~\ref{fig1}(b). The 1T structure corresponds to a monolayer that is truncated from the (111) surfaces of 3D phase. The 1T and 2H structures have a hexagonal unit cell with lattice constant $a$. The atoms are located at $A(0,0,0)$, $B(0,a/\sqrt{3},h)$, and $B(a/2,a/(2\sqrt{3}),-h)$ for 1T structure, and $A(0,0,0)$, $B(0,a/\sqrt{3},\pm h)$ for 2H structure, where $2h$ is the thickness of the monolayer. For the case of $A_2B$ systems, $A$ and $B$ atoms should be interchanged. The 1T$^{\prime}$ structure has a rectangular unit cell with lattice constants of $a$ and $b$. This is regarded as a distorted 1T structure, as $A$ atoms are dimerized along $y$ direction and $B$ atoms follow the movement of $A$ atoms. The 1T$^{\prime}$ structure may be buckled along the $z$ direction. The PCK structure is regarded as a monolayer truncated from the (110) surface. $A$ atom in the center of rectangular cell is displaced to $z$ direction, i.e., $A(a/2,b/2,\delta)$. Each $A$ atom is surrounded by four $B$ atoms in the same planes, $z = 0$ and $\delta$. 

\subsection{First-principles calculations}
We used Quantum ESPRESSO (QE) code \cite{qe} to perform density-functional theory (DFT) calculations. 
We used exchange-correlation energy functional within generalized-gradient approximation (GGA) parametrized by Perdew, Burke, and Ernzerhof (PBE) \cite{pbe}. The electron-ion interactions were treated by using ultrasoft pseudopotentials in pslibrary1.0.0 \cite{dalcorso}. The energy cutoff for wavefunction was set to be $E_{\rm cut}=\max(E_A,E_B)$ plus 20 Ry, where $E_j (j=A, B)$ is the suggested value for atom $j$, and the energy cutoff for charge density was set to be $10E_{\rm cut}$. Spin-polarized calculations were performed. A vacuum layer was set to be 15 \AA \ for 2D systems. Convergence thresholds for the total energy in the self-consistent field calculations was set to be $10^{-12}$ Ry, and those for the total energy and forces for structure optimization were set to be $10^{-5}$ Ry and $10^{-4}$ a.u., respectively. 

We first optimized the crystal structure by using smearing parameter of 0.015 Ry \cite{smearingMP} and $k$-point distance $\Delta k$ smaller than 0.1 \AA$^{-1}$. The lattice constant of $0.95a_{3D}/\sqrt{2}$ was assumed as an initial guess for 1T and 2H structures, where $a_{3D}$ is the lattice constant of the cubic phase and the factor of $0.95$ accounts for the in-plane contraction due to the lack of atoms along the out-of-plane direction. 
For 1T$^\prime$ structure,  an initial structure was prepared by referring to the crystal structure of WTe$_2$ monolayer. 
For PCK structure, $b/a\simeq \sqrt{2}$ and $\delta = 0.4a$ were assumed as an initial guess. Ferromagnetic phase was assumed in the initial spin configurations. For semiconducting systems, we optimized the structure again by using no smearing parameters and assuming $\Delta k \le 0.15 $ \AA$^{-1}$. 

The stability of $A_nB_m$ system is studied by calculating the formation energy
\begin{eqnarray}
 E_{\rm form}(A_nB_m) = \frac{\varepsilon_{\alpha}(A_nB_m) - n\varepsilon(A) - m\varepsilon(B)}{n+m},
\end{eqnarray}
where $n$ and $m$ are integers, $\varepsilon_{\alpha}(A_nB_m)$ is the total energy of $A_nB_m$ in the structure $\alpha$, and $\varepsilon(X)$ with $X=A, B$ is the total energy of element $X$. To obtain $\varepsilon(X)$, we extracted the ground state structure of $X$ from the Open Quantum Materials Database (OQMD) \cite{oqmd}, as done in the construction of Computational 2D Materials Database (C2DB) \cite{C2DB2021}. 
We optimized the structure and calculated $\varepsilon(X)$.  

For the bulk and the 2D system having negative formation energy and the lowest energy among four structures (1T, 2H, 1T$^\prime$, and PCK), we calculate the phonon dispersions within density-functional perturbation theory \cite{dfpt}. The long-range Coulomb correction is included in the force constant matrix for semiconductors. The Coulomb interaction in the $z$ direction is truncated for 2D systems \cite{iso2D}. The $8\times 8 \times 1$ $q$ grid is used for the 1T and 2H structures, $6\times 4\times 1$ for the 1T$^\prime$ and PCK structures, while $4\times 4 \times 4$ for the 3D structure. The variable for the acoustic sum rule is set to be ``crystal'' \cite{qe}. 

For the systems having NPR, first-principles molecular dynamics (MD) simulations are performed by using QE \cite{qe}. A $4\times 4\times 1$ supercell is assumed, the volume of the unit cell is fixed, and the ionic temperature is kept to 300 K by adapting the velocity scaling. The Newton's equation is integrated by using the Verlet algorithm with a time step of 1 fs.  

To investigate the ionic character of 2D $AB_2$ and $A_2B$ systems, we estimate the Madelung constant $M$ of 1T and 2H structures. The $M$ is then characterized by $h/a$ only. The electrostatic energy is calculated by using pymatgen code \cite{pymatgen}. 

We calculate the surface energy $\gamma$ for the (111) and (110) planes of fluorite-type materials. This is defined as
\begin{eqnarray}
 \gamma = \frac{E_{\rm slab} - nE_{\rm bulk}}{2A},
\end{eqnarray} 
where $E_{\rm slab}$ is the total energy of the slab including $n$ unit cells, $E_{\rm bulk}$ is the total energy of 3D bulk, and $A$ is the area of the surface. The factor of $1/2$ accounts for the presence of two surfaces on top and bottom sides of the slab. For the (111) and (110) surfaces, 7 and 9 layer-thick slabs (21 and 27 atoms) including a vacuum layer of 15 \AA \ are assumed and the atomic position of the middle (fourth and fifth) layer is fixed in the geometry optimization. The slab models are constructed by using atomic simulation environment (ASE) package \cite{ase}.

\section{Results and Discussion}
\subsection{Stability trend}
\label{sec:stability_trend}
Before studying the 2D systems, we calculated phonon dispersions of 53 compounds in the fluorite structure to check whether the methodology used in the present work correctly predicts the dynamical stability of 3D counterparts. 
We have found that HgF$_2$, PIr$_2$, ZrO$_2$, and HfO$_2$ are unstable within PBE. When Perdew-Zunger \cite{pz} and PBEsol \cite{pbesol} functionals are used, HgF$_2$ and PIr$_2$ are dynamically stable. 

Table \ref{table1}, \ref{table2}, and \ref{table3} lists the $E_{\rm form}$ of $AB_2$, $A_2B$ systems, and other ordered alloys, respectively, in the 1T, 2H, 1T$^{\prime}$, and PCK structures for 53 compounds. The $E_{\rm form}$ of bulk is also tabulated in these Tables. For $AB_2$ and $A_2B$ systems except for ScH$_2$, YH$_2$, TiF$_2$, PbF$_2$, and Li$_2$Te, 1T phase is the most stable 2D structure because 2H and PCK phases are higher in energy and 1T$^{\prime}$ phase is relaxed to 1T structure, as listed in Table \ref{table1} and \ref{table2}. The $E_{\rm form}$ of the 1T phase is larger than of bulk by less than 0.2 eV, while hydrides tend to be less stable. These 1T phases are dynamically stable. However, some of them show imaginary frequencies around $\Gamma$ point in the Brillouin zone. This is attributed to an instability against the flexural (out-of-plane) vibrations. Phonon dispersions for the lowest energy phase are provided in Supplemental Materials \cite{SM}. 

The stability preference of ordered alloys is different from $AB_2$ and $A_2B$ systems. As listed in Table \ref{table3}, the 1T$^{\prime}$ or PCK phases are preferred rather than 1T phase. However, many of them have positive $E_{\rm form}$, while they have negative $E_{\rm form}$ in the bulk. The systems having negative $E_{\rm form}$ are $X_2$Pt ($X=$ Al, Ga, In, and Sn) in the 1T$^\prime$ structure, PRh$_2$, Ga$_2$Au, and CoSi$_2$ in the PCK structure, and Al$_2$Pd in the 2H structure. These structures are all dynamically stable except for Sn$_2$Pt. The phonon dispersions are provided in Supplemental Materials \cite{SM}. 

\begin{table}\begin{center}\caption{Formation energy (eV/atom) of 3D bulk, 1T, 2H, 1T$^\prime$, and Puckered (PCK) structures for 19 $AB_2$ systems. When 1T structure is not a ground state among 2D phases, the value is underlined for the most stable structure. A hyphen indicates that no scf convergence or relaxed geometry are obtained. }
{\begin{tabular}{lrrrrr}\hline\hline
 \hspace{3mm} & bulk \hspace{3mm} &  1T \hspace{3mm} &  2H  \hspace{3mm} & 1T$^\prime$ \hspace{3mm} & PCK  \\
\hline
ScH$_{2}$ \hspace{3mm} & $ -0.67 $ \hspace{3mm} & $ -0.27 $ \hspace{3mm} & $ -0.20 $ \hspace{3mm} & $ -0.27 $ \hspace{3mm} & \underline{$ -0.29 $} \\
YH$_{2}$ \hspace{3mm} & $ -0.71 $ \hspace{3mm} & $ -0.25 $ \hspace{3mm} & $ -0.19 $ \hspace{3mm} & $ -0.25 $ \hspace{3mm} & \underline{$ -0.29 $} \\
TiH$_{2}$ \hspace{3mm} & $ -0.48 $ \hspace{3mm} & $ -0.16 $ \hspace{3mm} & $ -0.05 $ \hspace{3mm} & $ -0.16 $ \hspace{3mm} & $ -0.10 $ \\
ZrH$_{2}$ \hspace{3mm} & $ -0.56 $ \hspace{3mm} & $ -0.15 $ \hspace{3mm} & $ -0.02 $ \hspace{3mm} & $ -0.15 $ \hspace{3mm} & $ -0.10 $ \\
VH$_{2}$ \hspace{3mm} & $ -0.20 $ \hspace{3mm} & 0.02 \hspace{3mm} & $ 0.11 $ \hspace{3mm} & $ 0.02 $ \hspace{3mm} & $ 0.12 $ \\
NbH$_{2}$ \hspace{3mm} & $ -0.24 $ \hspace{3mm} & $ 0.09 $ \hspace{3mm} & $ 0.19 $ \hspace{3mm} & $ 0.09 $ \hspace{3mm} & $ 0.19 $ \\
TaH$_{2}$ \hspace{3mm} & $ -0.10 $ \hspace{3mm} & $ 0.16 $ \hspace{3mm} & $ 0.22 $ \hspace{3mm} & $ 0.16 $ \hspace{3mm} & $ 0.24 $ \\
CrH$_{2}$ \hspace{3mm} & $ 0.06 $ \hspace{3mm} & $ 0.17 $ \hspace{3mm} & $ 0.25 $ \hspace{3mm} & $ 0.17 $ \hspace{3mm} & - \\
CaF$_{2}$ \hspace{3mm} & $ -3.92 $ \hspace{3mm} & $ -3.79 $ \hspace{3mm} & $ -3.53 $ \hspace{3mm} & $ -3.79 $ \hspace{3mm} & $ -3.75 $ \\
SrF$_{2}$ \hspace{3mm} & $ -3.93 $ \hspace{3mm} & $ -3.75 $ \hspace{3mm} & $ -3.52 $ \hspace{3mm} & $ -3.75 $ \hspace{3mm} & $ -3.73 $ \\
BaF$_{2}$ \hspace{3mm} & $ -3.87 $ \hspace{3mm} & $ -3.69 $ \hspace{3mm} & $ -3.50 $ \hspace{3mm} & $ -3.69 $ \hspace{3mm} & $ -3.68 $ \\
TiF$_{2}$ \hspace{3mm} & $ -2.72 $ \hspace{3mm} & $ -2.62 $ \hspace{3mm} & \underline{$ -2.84 $} \hspace{3mm} & $ -2.80 $ \hspace{3mm} & - \\
CdF$_{2}$ \hspace{3mm} & $ -2.17 $ \hspace{3mm} & $ -2.09 $ \hspace{3mm} & $ -1.90 $ \hspace{3mm} & $ -2.09 $ \hspace{3mm} & $ -2.06 $ \\
HgF$_{2}$ \hspace{3mm} & $ -1.38 $ \hspace{3mm} & $ -1.33 $ \hspace{3mm} & $ -1.18 $ \hspace{3mm} & $ -1.33 $ \hspace{3mm} & $ -1.31 $ \\
PbF$_{2}$ \hspace{3mm} & $ -2.37 $ \hspace{3mm} & $ -2.23 $ \hspace{3mm} & $ -2.08 $ \hspace{3mm} & $ -2.26 $ \hspace{3mm} & \underline{$ -2.28 $} \\
SrCl$_{2}$ \hspace{3mm} & $ -2.50 $ \hspace{3mm} & $ -2.45 $ \hspace{3mm} & $ -2.31 $ \hspace{3mm} & $ -2.45 $ \hspace{3mm} & $ -2.42 $ \\
BaCl$_{2}$ \hspace{3mm} & $ -2.58 $ \hspace{3mm} & $ -2.47 $ \hspace{3mm} & $ -2.34 $ \hspace{3mm} & $ -2.47 $ \hspace{3mm} & $ -2.46 $ \\
ZrO$_{2}$ \hspace{3mm} & $ -3.38 $ \hspace{3mm} & $ -3.23 $ \hspace{3mm} & $ -2.68 $ \hspace{3mm} & $ -3.23 $ \hspace{3mm} & $ -3.19 $ \\
HfO$_{2}$ \hspace{3mm} & $ -3.58 $ \hspace{3mm} & $ -3.46 $ \hspace{3mm} & $ -2.86 $ \hspace{3mm} & $ -3.46 $ \hspace{3mm} & $ -3.40 $ \\
\hline
\end{tabular}}\label{table1}\end{center} \end{table}

\begin{table}\begin{center}\caption{Same as Table \ref{table1} but for 14 $A_2B$-type systems.}
{\begin{tabular}{lrrrrr}\hline\hline
 \hspace{3mm} & bulk \hspace{3mm} &  1T \hspace{3mm} &  2H  \hspace{3mm} & 1T$^\prime$ \hspace{3mm} & PCK  \\
\hline
Li$_{2}$O \hspace{3mm} & $ -1.86 $ \hspace{3mm} & $ -1.69 $ \hspace{3mm} & $ -1.38 $ \hspace{3mm} & $ -1.69 $ \hspace{3mm} & $ -1.65 $ \\
Li$_{2}$S \hspace{3mm} & $ -1.36 $ \hspace{3mm} & $ -1.16 $ \hspace{3mm} & $ -0.94 $ \hspace{3mm} & $ -1.16 $ \hspace{3mm} & $ -1.15 $ \\
Li$_{2}$Se \hspace{3mm} & $ -1.27 $ \hspace{3mm} & $ -1.08 $ \hspace{3mm} & $ -0.88 $ \hspace{3mm} & $ -1.08 $ \hspace{3mm} & $ -1.08 $ \\
Li$_{2}$Te \hspace{3mm} & $ -1.04 $ \hspace{3mm} & $ -0.86 $ \hspace{3mm} & $ -0.70 $ \hspace{3mm} & $ -0.85 $ \hspace{3mm} & \underline{$ -0.86 $} \\
Na$_{2}$O \hspace{3mm} & $ -1.22 $ \hspace{3mm} & $ -1.08 $ \hspace{3mm} & $ -0.86 $ \hspace{3mm} & $ -1.08 $ \hspace{3mm} & $ -1.04 $ \\
Na$_{2}$S \hspace{3mm} & $ -1.08 $ \hspace{3mm} & $ -0.90 $ \hspace{3mm} & $ -0.71 $ \hspace{3mm} & $ -0.90 $ \hspace{3mm} & $ -0.88 $ \\
Na$_{2}$Se \hspace{3mm} & $ -1.08 $ \hspace{3mm} & $ -0.89 $ \hspace{3mm} & $ -0.72 $ \hspace{3mm} & $ -0.89 $ \hspace{3mm} & $ -0.88 $ \\
Na$_{2}$Te \hspace{3mm} & $ -0.94 $ \hspace{3mm} & $ -0.76 $ \hspace{3mm} & $ -0.61 $ \hspace{3mm} & $ -0.76 $ \hspace{3mm} & $ -0.76 $ \\
K$_{2}$S \hspace{3mm} & $ -1.08 $ \hspace{3mm} & $ -0.93 $ \hspace{3mm} & $ -0.78 $ \hspace{3mm} & $ -0.93 $ \hspace{3mm} & $ -0.91 $ \\
K$_{2}$Se \hspace{3mm} & $ -1.11 $ \hspace{3mm} & $ -0.96 $ \hspace{3mm} & $ -0.82 $ \hspace{3mm} & $ -0.96 $ \hspace{3mm} & $ -0.94 $ \\
K$_{2}$Te \hspace{3mm} & $ -1.03 $ \hspace{3mm} & $ -0.88 $ \hspace{3mm} & $ -0.74 $ \hspace{3mm} & $ -0.88 $ \hspace{3mm} & $ -0.86 $ \\
Rb$_{2}$S \hspace{3mm} & $ -1.02 $ \hspace{3mm} & $ -0.90 $ \hspace{3mm} & $ -0.76 $ \hspace{3mm} & $ -0.90 $ \hspace{3mm} & $ -0.87 $ \\
Rb$_{2}$Se \hspace{3mm} & $ -1.07 $ \hspace{3mm} & $ -0.94 $ \hspace{3mm} & $ -0.81 $ \hspace{3mm} & $ -0.94 $ \hspace{3mm} & $ -0.91 $ \\
Rb$_{2}$Te \hspace{3mm} & $ -1.01 $ \hspace{3mm} & $ -0.87 $ \hspace{3mm} & $ -0.74 $ \hspace{3mm} & $ -0.87 $ \hspace{3mm} & $ -0.85 $ \\
\hline
\end{tabular}}\label{table2}\end{center} \end{table}

\begin{table}\begin{center}\caption{Same as Table \ref{table1} but for 20 ordered alloys.}
{\begin{tabular}{lrrrrr}\hline\hline
\hspace{3mm} & bulk \hspace{3mm} &  1T \hspace{3mm} &  2H  \hspace{3mm} & 1T$^\prime$ \hspace{3mm} & PCK  \\
\hline
Be$_{2}$C \hspace{3mm} & $ -0.24 $ \hspace{3mm} & $ 0.21 $ \hspace{3mm} & $ 0.82 $ \hspace{3mm} & $ 0.56 $ \hspace{3mm} & $ 0.21 $ \\
Be$_{2}$B \hspace{3mm} & $ 0.04 $ \hspace{3mm} & $ 0.45 $ \hspace{3mm} & $ 0.56 $ \hspace{3mm} & $ 0.53 $ \hspace{3mm} & $ 0.45 $ \\
Mg$_{2}$Si \hspace{3mm} & $ -0.16 $ \hspace{3mm} & $ 0.30 $ \hspace{3mm} & $ 0.34 $ \hspace{3mm} & $ 0.28 $ \hspace{3mm} & \underline{$ 0.27 $} \\
Mg$_{2}$Ge \hspace{3mm} & $ -0.27 $ \hspace{3mm} & $ 0.17 $ \hspace{3mm} & $ 0.20 $ \hspace{3mm} & $ 0.19 $ \hspace{3mm} & \underline{$ 0.13 $} \\
Mg$_{2}$Sn \hspace{3mm} & $ -0.20 $ \hspace{3mm} & $ 0.21 $ \hspace{3mm} & $ 0.49 $ \hspace{3mm} & \underline{$ 0.15 $} \hspace{3mm} & $ 0.16 $ \\
Mg$_{2}$Pb \hspace{3mm} & $ -0.07 $ \hspace{3mm} & $ 0.32 $ \hspace{3mm} & $ 0.55 $ \hspace{3mm} & $ 0.25 $ \hspace{3mm} & \underline{$ 0.24 $} \\
AsRh$_{2}$ \hspace{3mm} & $ -0.42 $ \hspace{3mm} & $ 0.44 $ \hspace{3mm} & $ 0.40 $ \hspace{3mm} & $ 0.29 $ \hspace{3mm} & \underline{$ 0.25 $} \\
PRh$_{2}$ \hspace{3mm} & $ -0.75 $ \hspace{3mm} & $ 0.06 $ \hspace{3mm} & $ 0.06 $ \hspace{3mm} & $ -0.01 $ \hspace{3mm} & \underline{$ -0.06 $} \\
PIr$_{2}$ \hspace{3mm} & $ -0.37 $ \hspace{3mm} & $ 0.75 $ \hspace{3mm} & $ 0.46 $ \hspace{3mm} & \underline{$ 0.40 $} \hspace{3mm} & $ 0.45 $ \\
Sn$_{2}$Ir \hspace{3mm} & $ -0.25 $ \hspace{3mm} & $ 0.22 $ \hspace{3mm} & $ 0.23 $ \hspace{3mm} & \underline{$ 0.06 $} \hspace{3mm} & $ 0.21 $ \\
Al$_{2}$Pt \hspace{3mm} & $ -0.89 $ \hspace{3mm} & $ -0.22 $ \hspace{3mm} & $ -0.05 $ \hspace{3mm} & \underline{$ -0.29 $} \hspace{3mm} & $ -0.22 $ \\
Ga$_{2}$Pt \hspace{3mm} & $ -0.58 $ \hspace{3mm} & $ -0.12 $ \hspace{3mm} & $ -0.02 $ \hspace{3mm} & \underline{$ -0.23 $} \hspace{3mm} & $ -0.14 $ \\
In$_{2}$Pt \hspace{3mm} & $ -0.48 $ \hspace{3mm} & $ -0.02 $ \hspace{3mm} & $ 0.01 $ \hspace{3mm} & \underline{$ -0.13 $} \hspace{3mm} & $ -0.08 $ \\
Sn$_{2}$Pt \hspace{3mm} & $ -0.45 $ \hspace{3mm} & $ -0.08 $ \hspace{3mm} & $ -0.07 $ \hspace{3mm} & \underline{$ -0.18 $} \hspace{3mm} & $ -0.13 $ \\
Al$_{2}$Au \hspace{3mm} & $ -0.44 $ \hspace{3mm} & $ 0.09 $ \hspace{3mm} & $ 0.13 $ \hspace{3mm} & $ 0.11 $ \hspace{3mm} & $ 0.10 $ \\
Ga$_{2}$Au \hspace{3mm} & $ -0.24 $ \hspace{3mm} & $ 0.10 $ \hspace{3mm} & $ 0.06 $ \hspace{3mm} & $ 0.04 $ \hspace{3mm} & \underline{$ -0.01 $} \\
In$_{2}$Au \hspace{3mm} & $ -0.25 $ \hspace{3mm} & $ 0.12 $ \hspace{3mm} & $ 0.07 $ \hspace{3mm} & $ 0.03 $ \hspace{3mm} & \underline{$ 0.01 $} \\
Si$_{2}$Ni \hspace{3mm} & $ -0.36 $ \hspace{3mm} & $ 0.21 $ \hspace{3mm} & $ 0.15 $ \hspace{3mm} & $ 0.13 $ \hspace{3mm} & \underline{$ 0.01 $} \\
CoSi$_{2}$ \hspace{3mm} & $ -0.53 $ \hspace{3mm} & $ 0.18 $ \hspace{3mm} & $ 0.15 $ \hspace{3mm} & $ 0.08 $ \hspace{3mm} & \underline{$ -0.06 $} \\
Al$_{2}$Pd \hspace{3mm} & $ -0.61 $ \hspace{3mm} & $ -0.01 $ \hspace{3mm} & \underline{$ -0.15 $} \hspace{3mm} & $ -0.13 $ \hspace{3mm} & $ -0.01 $ \\
\hline
\end{tabular}}\label{table3}\end{center} \end{table}

\begin{figure}
\center\includegraphics[scale=0.4]{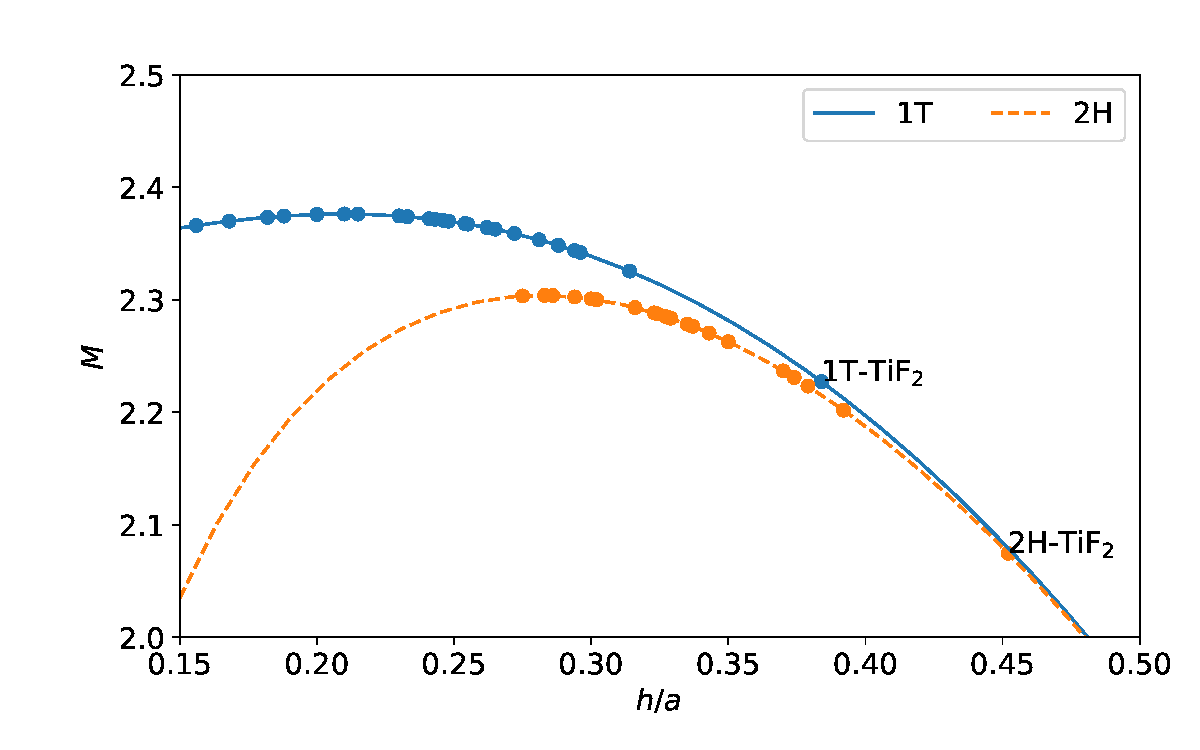}
\caption{The Madelung constant $M$ of 1T (solid) and 2H (dashed) structures as a function of $h/a$. The circles indicate $M$ of 25 systems listed in Table \ref{table1} and \ref{table2} except for hydrides. TiF$_2$ has a large $h/a$, and $M$ is small compared to the other compounds. }
\label{fig2} 
\end{figure}

We next calculate the Madelung constant $M$ to study the ionic character of $AB_2$ and $A_2B$ systems. The hydrides are excluded due to the metallic band structure. We assume the oxidation states of $B^-$ for halogens (F and Cl) and and $B^{2-}$ for chalcogens (O, S, Se, and Te). Figure \ref{fig2} plots $M$ as a function of $h/a$ for $AB_2$ and $A_2B$ systems. The 1T structures have $h/a\in [0.15,0.32]$ and $M>2.3$, whereas the 2H structures have $M<2.3$. Therefore, the ionic character of 1T structure is more significant than of 2H structure. 

Note that TiF$_2$ monolayer has a large value of $h/a=0.48$ and prefers 2H structure. It is interesting that 2H TiF$_2$ is more stable than the 3D bulk \cite{morita} (see Table \ref{table1}). The 2H structure has an indirect band gap of 1.3 eV (PBE) from K to $\Gamma$ in the Brillouin zone, while the 1T structure is metallic. This relationship between the crystal structure and electronic property is similar to that in MoS$_2$ monolayer \cite{magneMoS22018}. Such a structure-property relationship reflects the electronic configuration of transition metals. For Ti atom [Ar]$(3d)^2(4s)^2$, and a possible oxidation state is Ti$^{2+}$. For Mo atom [Kr]$(4d)^5(5s)^1$, and a possible oxidation state is Mo$^{4+}$ by assuming S$^{2-}$. The remaining two electrons in the Ti and Mo atoms should be occupied into $d$ energy levels. As the 2H structure has the D$_{3h}$ symmetry, $d_{z^2}$ state becomes the lowest energy state. By considering the spin degeneracy, the $d_{z^2}$ state is completely filled, while the other $d$ states are empty, resulting in a finite band gap. On the other hand, the 1T structure has the O$_h$ symmetry, and $d_{xy}$, $d_{yz}$, and $d_{zx}$ states become stable in energy. These states are partially filled with two electrons, resulting in a metallic phase. 

As we have obtained stable structures of fluorite-type materials in the monolayer limit, we next explore physical properties of PCK structures below. 

\subsection{Negative Poisson's ratio}
\begin{figure*}
\center\includegraphics[scale=0.55]{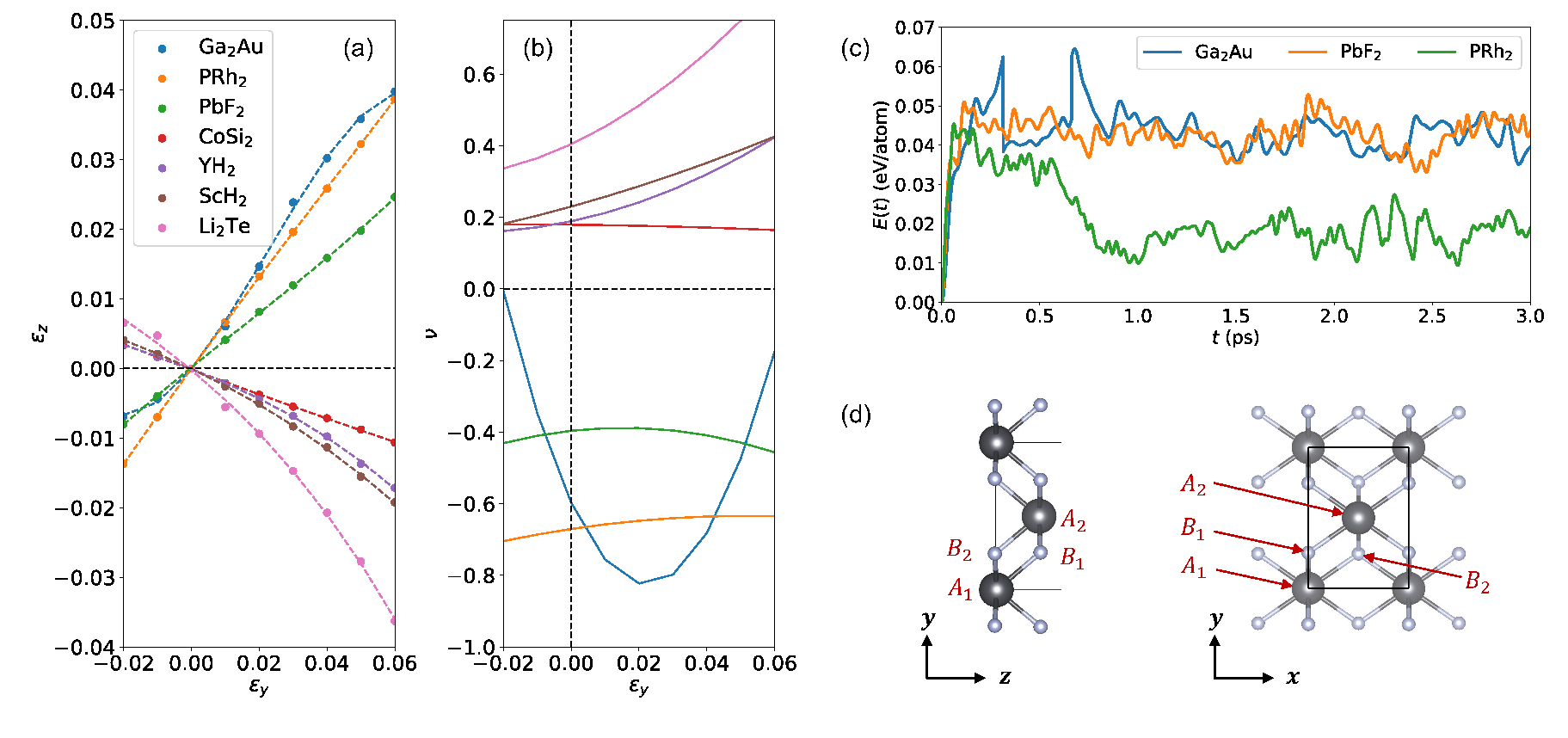}
\caption{The $\varepsilon_y$-dependence of (a) the out-of-plane strain $\varepsilon_z$ and (b) the Poisson's ratio $\nu=-\partial \varepsilon_z/\partial \varepsilon_y$ for the PCK structures. The third order polynomial is fitted to points ($\varepsilon_y,\varepsilon_z$), and $\nu$ is obtained from the derivative of the polynomial with respect to $\varepsilon_y$. (c) MD runs at 300 K for Ga$_2$Au, PbF$_2$, and PRh$_2$ monolayers in the PCK phase. A $4\times 4 \times 1$ supercell is assumed (96 atoms). PRh$_2$ monolayer becomes unstable after 1 ps. (d) $A_1, A_2, B_1$, and $B_2$ atoms for the PCK-structured $AB_2$ monolayer (see Sec.~\ref{sec:ana}). }
\label{fig3} 
\end{figure*}

We study the elastic property of the PCK structures: ScH$_2$, YH$_2$, PbF$_2$, Li$_2$Te, PRh$_2$, Ga$_2$Au, and CoSi$_2$.
The PCK structure is stretched along the $y$ direction. We define the out-of-plane strain as $\varepsilon_z = (d-d_0)/d_0$, where $d$ and $d_0$ are the thickness of the monolayer with and without the strain, respectively.
Figure \ref{fig3}(a) shows $\varepsilon_z$ as a function of $\varepsilon_y$. The atom position and the lattice constant $a$ are optimized, while the lattice constant $b$ is fixed. 
Interestingly, when $\varepsilon_y$ is increased, $\varepsilon_z$ also increases for Ga$_2$Au, PbF$_2$, and PRh$_2$ monolayers. 
This indicates an NPR in the out-of-plane direction, $\nu=-\partial \varepsilon_z/\partial \varepsilon_y<0$. 
As shown in Fig.~\ref{fig3}(b), $\nu$ takes a minimum value of $-0.8$ at $\varepsilon_y=0.02$ in Ga$_2$Au monolayer. 
First-principles MD simulations show that they are thermodynamically stable at 300 K except for PRh$_2$, as shown in Fig.~\ref{fig3}(c). 
For the case of 2D semiconductors (PbF$_2$ and Li$_2$Te), we plotted the potential energy and $\varepsilon_z$ as functions of $(a-a_0)/a_0$ and $(b-b_0)/b_0$, where $a_0$ and $b_0$ are the lattice constants in equilibrium. The NPR is observed only when the PbF$_2$ monolayer is stretched along the $y$ direction. This is provided in the Supplemental Material \cite{SM}. 

We have investigated whether the NPR also appears for thicker PbF$_2$ films. The PCK structure is constructed by stacking two PbF$_2$ layers [see Fig.~\ref{fig1}(a)]. Therefore, we have constructed three (9 atoms) and four (12 atom) PbF$_2$ layers. The strain is increased from $\varepsilon_y=0$ to 0.06. However, no NPR is observed in these systems because the thickness monotonically decreases from 4.017 and 6.134 \AA \ to 3.947 and 6.056 \AA \ for three and four layers, respectively.

\subsubsection{Analytical model}
\label{sec:ana}
We discuss an origin of the NPR based on an analytical model. 
As shown in Fig.~\ref{fig3}(d), the atomic positions of the PCK structure are denoted as $A_1(0,0,0)$, $A_2(a/2,b/2,z_A)$, $B_1(0,b/4,z_B)$, and $B_2(a/2,b/4,z_A-z_B)$, where $z_A$ and $z_B$ are the $z$ coordinates of $A_2$ and $B_1$ atoms, respectively, and $B_2$ atom has $z=z_A-z_B$. The thickness is given by $d=2z_B - z_A$ when $z_A<z_B$.  
When the monolayer is elongated along the $y$ direction ($\varepsilon_y>0$), it will be contracted along the $x$ direction ($\varepsilon_x<0$), and the $A_2$ and $B_1$ atoms will approach the $z=0$ plane ($\varepsilon_{z}^{A}<0$ and $\varepsilon_{z}^{B}<0$). Then, the $z$ coordinate of $B_2$ atom becomes negatively large, which results in an increase in $d$, {\it i.e.}, the NPR. 
 
To show this, we assume that the bond lengths, $l(A_1A_2), l(A_1B_1)$, and $l(A_1B_2)$, are not changed under the strain.
This is a good approximation for the PCK structures: For PbF$_2$, $l(A_1A_2)=4.02\rightarrow 4.06$ \AA, $l(A_1B_1)=2.33\rightarrow 2.35$ \AA, and $l(A_1B_2)=2.57\rightarrow 2.58$ \AA \ when $\varepsilon_y=0.06$ is applied. 
The structural parameters are changed from $a, b,$ $z_A$, and $z_B$ to $a'=a(1+\varepsilon_x)$, $b'=b(1+\varepsilon_y)$, $z_{A}' = z_A(1+\varepsilon_{z}^{A})$, and $z_{B}' = z_B(1+\varepsilon_{z}^{B})$, respectively. When $l(A_1A_2)$ is equal to $l'(A_1A_2)$ under the strain, the following equation should be satisfied: 
\begin{eqnarray}
 a^2 \varepsilon_x + b^2 \varepsilon_y + 4z_{A}^2  \varepsilon_{z}^{A} = 0,
\label{eq:A1}
\end{eqnarray}
where the first order of strains is considered. From the conditions $l(A_1B_1) = l'(A_1B_1)$ and $l(A_1B_2)=l'(A_1B_2)$, we obtain the relationships
\begin{eqnarray}
 b^2 \varepsilon_y + 16z_{B}^2  \varepsilon_{z}^{B} = 0
\label{eq:A2}
\end{eqnarray}
and
\begin{eqnarray}
 4a^2 \varepsilon_x + b^2 \varepsilon_y + 16(z_{A}- z_{B}) (z_{A}\varepsilon_{z}^{A} - z_{B}\varepsilon_{z}^{B})  = 0.
\label{eq:A3}
\end{eqnarray}
Eq.~(\ref{eq:A2}) indicates that the $z$ coordinate of $B_1$ atom decreases to $z_B(1+\varepsilon_{z}^{B})$ because $\varepsilon_{z}^{B}<0$ when $\varepsilon_y >0$. From Eqs.~(\ref{eq:A1})-(\ref{eq:A3}), we obtain
\begin{eqnarray}
 \varepsilon_{z}^{A} 
 = - \frac{b^2}{16z_Az_B} \left(4- \frac{z_A}{z_B}\right) \varepsilon_y.
\label{eq:A4}
\end{eqnarray}
This also predicts $\varepsilon_{z}^{A}<0$ because $z_A \simeq z_B$. 
The $\varepsilon_{z}^{A}$ is negatively larger than $\varepsilon_{z}^{B}$ by a factor of three. 
Therefore, the thickness, expressed by
\begin{eqnarray}
d = d_0 +  \frac{b^2}{8z_B} \left(1 - \frac{z_A}{2z_{B}} \right) \varepsilon_y
\label{eq:A5}
\end{eqnarray}
with $d_0=2z_B-z_A$, increases with $\varepsilon_y$.
This results in $\nu = - \partial (d/d_0)/\partial \varepsilon_y<0$. 
For an unrelaxed PCK structure ($b=\sqrt{2}a$ and $d_0=z_A=z_B=a/2$), $\nu$ is exactly equal to $-1/2$. 

In our model, $\nu<0$ always holds. In realistic systems, the bond lengths are changed within a few percent under the strains. In addition, $B_1$ and $B_2$ atoms have different $y$ coordinates. These affects the $\varepsilon_y$-dependence of $\nu$. 

\subsubsection{Relationship to the surface energy}
Our analytical model suggests that the NPR is inherently present in the PCK structure.  
As the PCK structure is truncated from the (110) surface, it is interesting to study whether the monolayers having NPR are identified by surface energy calculations. 

\begin{figure*}
\center
\includegraphics[scale=0.65]{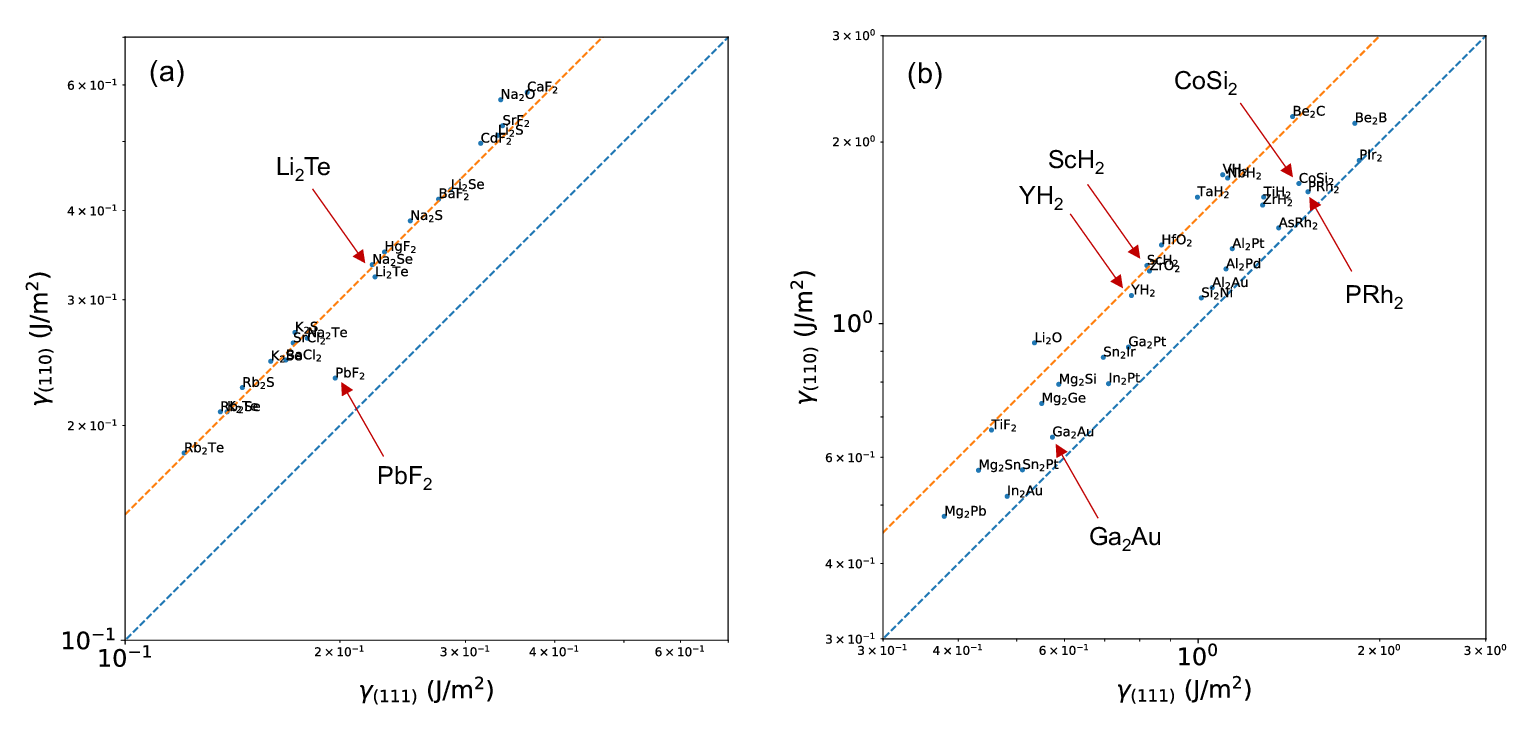}
\caption{Correlation between $\gamma_{(111)}$ and $\gamma_{(110)}$. The dashed lines indicate $\gamma_{(110)}=\gamma_{(111)}$ (blue) and $\gamma_{(110)}=1.5\gamma_{(111)}$ (orange). } \label{fig4} 
\end{figure*}

Figure \ref{fig4} shows a correlation of the surface energies between the (111) and (110) surfaces.
The $\gamma$ for $AB_2$ and $A_2B$ systems is plotted in Fig.~\ref{fig4}(a), while that for the other ordered alloys is plotted in Fig.~\ref{fig4}(b). The calculated data for ZrO$_2$, HfO$_2$, Li$_2$O, and Be$_2$O are plotted in Fig.~\ref{fig4}(b) due to the large $\gamma$. CrH$_2$ is excluded because the scf calculation assuming ferromagnetic phase is not converged. For all systems, the (111) surface is more stable. This is consistent with previous calculations on CaF$_2$, SrF$_2$, and BaF$_2$ \cite{bebelis2017}. As shown in Fig.~\ref{fig4}(a), a linear relationship of $\gamma_{(110)}\simeq 1.5\gamma_{(111)}$ holds for $AB_2$ and $A_2B$ systems. For the other ordered alloys in the high energy regime ($\gamma_{(111)}\gtrsim 0.4$ J/m$^2$), the relation $\gamma_{(111)}<\gamma_{(110)}\lesssim 1.5\gamma_{(111)}$ is satisfied, as shown in Fig.~\ref{fig4}(b). It is noteworthy that PbF$_2$, Ga$_2$Au, and PRh$_2$ having NPR exhibit a large deviation from the linear relationship of $\gamma_{(110)}\simeq 1.5\gamma_{(111)}$. 

\section{Conclusion}
We have studied 53 fluorite-type materials in the monolayer limit by performing first principles calculations. Most of $AB_2$ and $A_2B$ systems form the 1T structure. The puckered structure can exhibit a negative Poisson's ratio (NPR) in the out-of-plane direction. 
An analytical model is developed to explain the NPR. 
The puckered structures having NPR (PbF$_2$, Ga$_2$Au, and PRh$_2$) are identified by the energy difference between the (111) and (110) surfaces. 

In the present work, we have focused on non-layered fluorite-type materials to explore novel 2D materials. 
The 1T and puckered structures are truncated from the (111) and (110) surfaces, respectively. 
A similar strategy, truncation from low-index surfaces of non-layered materials, will be useful to expand the family of 2D materials. 

\begin{acknowledgments}
This work was supported by JSPS KAKENHI (Grant No. 21K04628). Calculations were done using the facilities of the Supercomputer Center, the Institute for Solid State Physics, the University of Tokyo, and the Supercomputer ``MASAMUNE-IMR'' at Center for Computational Materials Science, Institute for Materials Research, Tohoku University.
\end{acknowledgments}




\end{document}